\documentclass[aps,showpacs,preprintnumbers,amsmath,epsf,amssymb,epsfig,nofootinbib]{revtex4}
\usepackage{bm}
\begin{document}

\title{Passing to an effective 4D phantom cosmology from 5D vacuum theory of gravity}
\author{$^{1}$Jos\'e Edgar Madriz Aguilar\footnote{ E-mail address:
jemadriz@fisica.ufpb.br} and $^{2,3}$Mauricio
Bellini\footnote{E-mail address: mbellini@mdp.edu.ar}}

\address{$^1$  Departamento de F\'{\i}sica, Universidade
Federal da Para\'{\i}ba,\\
Caixa Postal 5008, 58059-970 Jo\~{a}o Pessoa, Pb, Brazil.\\
$^{2}$Departamento de
F\'{\i}sica, Facultad de Ciencias Exactas y Naturales, Universidad
Nacional de Mar del Plata, Funes 3350, (7600) Mar del Plata,
Argentina.\\
$^3$ Consejo Nacional de Investigaciones
Cient\'{\i}ficas y T\'ecnicas (CONICET).
}

\begin{abstract}
Starting from a five-dimensional (5D) vacuum theory of gravity where the extra coordinate is considered as noncompact, we 
investigate the possibility of inducing four-dimensional
(4D) phantom scenarios by applying form-invariance symmetry transformations. In
particular we obtain phantom scenarios for two cosmological frameworks. In the first framework we deal with an induced 4D 
de-Sitter expansion and in the second one a 4D induced model where the expansion of the universe is dominated by a
decreasing cosmological parameter $\Lambda(t)$ is discussed.
\end{abstract}

\maketitle

\section{Introduction}

There is observational evidence that the universe is spatially
flat, low matter density and currently undergoing to a period of
accelerating expansion. The implications for cosmology should be
that the cosmological fluid is dominated by some sort of fantastic
energy density, which has negative pressure and play an important
role today. Experimental evidence suggests that the present values
of the dark energy and matter components, in terms of the critical
density, are approximately $\Omega_{\chi}\simeq 0.7$ and
$\Omega_{M} \simeq 0.3$\cite{turner}. If one makes the most
conservative assumption that $\Omega_\chi$ corresponds to a
cosmological parameter which is constant, hence the equation of
state is given by a constant $\omega _{\chi}= {\rm P}/\rho = -1$,
which describes a vacuum dominated
universe.\\

On the other hand, in the last years there has been an intense
interest in finding exact solutions of the 5D field equations of the by some called noncompact Kaluza-Klein theories (NKK), on 
which the fifth coordinate (in our case $\psi$) is considered as
noncompact. Unlike the usual KK theory in which a cyclic symmetry
related to the extra dimension is assumed, the new approach
removes the cyclic condition on the extra dimension retaining
derivatives of the metric with respect to the extra coordinate.
This induces non-trivial matter on the hypersurfaces with
constant-$\psi$ and other non trivial frames. This theory
reproduces and extends known solutions of the Einstein field
equations in 4D\cite{rom} as well as quantum field theory in
curved
Lorentzian metrics\cite{nuestros}. On that framework several cosmological models can be obtained and being based on their 
particular observational predictions, it could be established a kind of equivalence between them. \\

An alternative equivalence relation among several cosmological models has been pondered in terms of the form invariance of their 
dynamical equations under a group of symmetry transformations that preserve the form of the Einstein field equations. Claiming 
invariance under this transformations, known as form-invariance symmetry transformations, inflationary cosmological scenarios from 
non-inflationary ones can be derived \cite{Chimento1,Chimento2}. One even more interesting feature of form-invariance 
transformations is that they can be used also to construct phantom cosmologies from cosmologies where the expansion of the 
universe is governed by a real scalar field $\phi$ \cite{Chimento3}. A form-invariance transformation relates the scale factor 
$a$, the Hubble parameter $H$, the energy density $\rho$ and the pressure $P$ of the initial cosmological model, with the 
cosmological parameters $\bar{a}$, $\bar{H}$, $\bar{\rho}$ and $\bar{P}$ of the so called transformed model. Such a transformation 
leaves invariant the form of the Friedmann dynamical equations of both models. As it was shown in \cite{Chimento3}, under the 
transformation $\bar{\rho}=n^{2}\rho$ with $n$ being a constant, the scalar field and its potential transform as
\begin{eqnarray}
\dot{\bar\phi}^{2}&=&n\dot{\phi}^{2},\nonumber \\
\bar{V}&=&n^{2}\left(\frac{1}{2}\dot{\phi}^{2}+V\right)-\frac{n}{2}\dot{\phi}^{2},\nonumber
\end{eqnarray}
where $\bar{V}$ is denoting the phantom potential. Thus it can be easily proved that when $n=\pm 1$, the $(+)$ branch corresponds 
to an identical transformation whereas the $(-)$ branch leads to a phantom cosmology. In that case the previous expressions become
\begin{eqnarray}
\dot{\bar\phi}^{2}&=&-\dot{\phi}^{2},\nonumber\\
\bar{V}(\bar\phi)&=&\dot{\phi}^{2}+V(\phi). \nonumber
\end{eqnarray}
Clearly the energy density of the transformed model $\bar\rho 
=(1/2)\dot{\bar\phi}^{2}+\bar{V}(\bar\phi)=-(1/2)\dot{\phi}^{2}+\bar{V}(\bar\phi)$ corresponds to a phantom scalar field 
$\bar\phi$ with the relation $\bar\phi =i\phi$ being valid.\\

In this letter we investigate the possibility of using form-invariance symmetry transformations to obtain phantom scenarios from a 
class of 4D universes induced from 5D vacuum which are vacuum dominated $({\rm P}=-\rho)$ and whose expansion is driven by a 
scalar field. The mechanism of dimensional reduction here adopted consists in establishing a foliation on the extra dimension 
inducing this way a 4D universe on every hypersurface. In section II we pass to a phantom scenario from a 4D induced universe with 
a de-Sitter expansion. Along section III the phantom scenario obtained is from a 4D induced model where the expansion of the 
universe is dominated by a decreasing cosmological parameter $\Lambda (t)$. Finally section IV is left for some final comments.

\section{Inducing a 4D phantom cosmological scenario on a de Sitter expansion}

Along the preset section we derive an induced 4D cosmological scenario by considering a deSitter expansion on the 4D hypersurfaces 
from a 5D space-time in geometrical vacuum. We start by introducing the 5D line element
\begin{equation}\label{y1}
dS^{2}= \psi ^{2}dN^{2}-\psi^{2}e^{2N}dr^{2}-d\psi^{2},
\end{equation}
where in this case $(N,r)$ are dimensionless coordinates being $dr^{2}=dX^{2}+dY^{2}+dZ^{2}$ with $(X,Y,Z)$ space-like coordinates 
and $N$ the time-like coordinate. The fifth dimension is again denoted by $\psi$ and has spatial units.  The metric in (\ref{y1}) 
has been extensively used in \cite{metrica,dina5d} to study cosmological scenarios induced from 5D vacuum due to its property of 
being Riemann-flat.\\

On the geometrical background (\ref{y1}) we consider the action
\begin{equation}\label{y2}
{\cal S}=\int d^{4}x\,d\psi \sqrt{-\,^{(5)}\!g}\left[\frac{^{(5)}\!R}{16\pi G}-\frac{1}{\psi 
^2}\left(\frac{1}{2}\,\overset{\star}{\varphi}^{2}-\frac{1}{2}\,e^{-2N}(\nabla 
_{r}\varphi)^{2}\right)+\frac{1}{2}\left(\frac{\partial\varphi}{\partial\psi}\right)^{2}\right],
\end{equation}
where the star $(\star)$ is denoting derivative respect to $N$ and $\varphi$ is a 5D free scalar field minimally coupled to 
gravity consistent with a 5D space-time in apparent vacuum
\cite{vacio}. Notice that the $\varphi$-Lagrangian in (\ref{y2})
is purely kinetic: ${\cal L}_{\varphi} \sim (1/2)\sqrt{-^{(5)} g}\,g^{AB} \varphi_{,A} \varphi_{,B}$. This is due to the fact 
that we are considering
an apparent vacuum in absence of any interactions. In this sense,
the scalar field $\varphi$ can be considered as a massless test
field on the 5D vacuum. The 5D dynamics of $\varphi$ obeys the
equation
\begin{equation}\label{y3}
\overset{\star\star}{\varphi}+3\overset{\star}{\varphi}-e^{-2N}\nabla _{r}^{2}\varphi-\left[4\psi\frac{\partial\varphi}{\partial 
\psi}+\psi^{2}\frac{\partial ^{2}\varphi}{\partial\psi ^2}\right]=0,
\end{equation}
which assuming separability of the scalar field $\varphi (N,\vec{r},\psi)=F(N,\vec{r})\Gamma(\psi)$ we note that the part 
depending of the fifth coordinate $\Gamma(\psi)$ obeys also the equation (\ref{p4}) and therefore $\Gamma(\psi)$ is given by 
expression (\ref{sol}).\\

In order to have the dynamics in terms of physical coordinates we use the coordinate transformation
\begin{equation}\label{y4}
t=\frac{N}{H_{0}},\qquad R=\frac{r}{H_0},\qquad \psi =\psi,
\end{equation}
where $t$ is the cosmic time and $R$ has units of length. Now applying the transformation (\ref{y4}), on the 4D hypersurface $\psi 
=H_{0}^{-1}$ the line element (\ref{y1}) becomes
\begin{equation}\label{y5}
ds^{2}=dt^{2}-e^{2H_{0}t}dR^{2},
\end{equation}
being $H_{0}$ the constant Hubble parameter. On this 4D hypersurface the action (\ref{y2}) reads
\begin{equation}\label{mod1}
\left.{\cal S}\right|_{(\psi=H_{0}^{-1})}=\int d^{4}x\,\sqrt{-\,^{(4)}\!g}\left[\frac{^{(4)}\!R}{16\pi 
G}-\left(\frac{1}{2}\dot{\varphi}^{2}-\frac{1}{2}\,e^{-2H_{0}t}(\nabla 
_{R}\varphi)^{2}-\frac{1}{2}\left(\frac{\partial\varphi}{\partial\psi}\right)^{2}_{\psi =H_{0}^{-1}}\right)\right],
\end{equation}
where now $\varphi=\varphi (t,\vec{R})$ and $^{(4)}\!R$ is the 4D Ricci scalar calculated with the induced metric (\ref{y5}). The 
scalar-field dynamics is governed by
\begin{equation}\label{y6}
\ddot{\varphi}+3H_{0}\dot{\varphi}-e^{-2H_{0}t}\nabla _{R}^{2}\varphi-H_{0}^{2}\left.\left[4\psi\frac{\partial\varphi}{\partial 
\psi}+\psi^{2}\frac{\partial ^{2}\varphi}{\partial\psi ^2}\right]\right|_{\psi =H_{0}^{-1}}=0,
\end{equation}
with $\varphi(t,\vec{R})=\varphi(t,\vec{R},\psi =H_{0}^{-1})$. In
order to describe a massive scalar field with an effective
quadratic $\varphi$-potential on an effective $\psi=H^{-1}_0$
hypersurface in the effective 4D action (\ref{mod1}), we shall
consider the particular case of a separable scalar field $\varphi
(t,\vec{r},\psi)=\bar\varphi(t,\vec{r})\Gamma(\psi)$. The equation
of motion for $\Gamma(\psi)$ results
\begin{equation}\label{p4}
\psi ^{2}\frac{d^{2}\Gamma}{d\psi
^2}+4\psi\frac{d\Gamma}{d\psi}-m^{2}\Gamma=0,
\end{equation}
being $m^{2} $ a separation constant. The general solution for
this equation is
\begin{equation}\label{sol}
\Gamma(\psi) = A \  \psi^{\sigma^+} + B \  \psi^{\sigma^-},
\end{equation}
being $A$ and $B$ integration constants and
\begin{equation}\label{sigma}
\sigma^{\pm} = -\frac{3}{2} \pm \frac{1}{2} \sqrt{9+4 m^2}.
\end{equation}
According to the action (\ref{mod1}) the induced 4D scalar potential, here called effective 4D
potential $V_{eff}(\varphi)$ has the form
\begin{equation}\label{y7}
V_{eff}(\bar{\varphi})=-\frac{1}{2}\,g^{\psi\psi}\left.\left(\frac{\partial\varphi}{\partial \psi}\right)^{2}\right|_{\psi 
=H_{0}^{-1}}=\frac{H_{0}^{2}}{2}\frac{\left[A\sigma ^{+}H_{0}^{\sigma ^{-}}+B\sigma 
^{-}H_{0}^{\sigma^{+}}\right]^2}{\left[AH_{0}^{\sigma
^{-}}+BH_{0}^{\sigma^{+}}\right]^2}\,\bar{\varphi}^{2}(t,\vec{R}),
\end{equation}
where $\sigma^{\pm}$ are given by equation (\ref{sigma}). This 4D effective potential is considered as classical.
As it is usually done in cosmology let us considering a global homogeneous scalar field defined by the spatial average  
$\left<\bar{\varphi}(t,\vec{R})\right>=\phi(t)$. This way the
dynamical equation (\ref{y6}) now in terms of the field $\phi(t)$
reads
\begin{equation}\label{y8}
\ddot{\phi}+3H_{0}\dot{\phi}+\frac{dV_{eff}(\phi)}{d\phi}=0,\,\rightarrow\,\ddot{\phi}+3H_{0}\dot{\phi}+M_{0}^{2}\phi=0
\end{equation}
where using (\ref{sol}) we have expressed the constant parameter $M_{0}$ by
\begin{equation}\label{y9}
M_{0}^{2}=H_{0}^{2}\left[\frac{A\sigma^{+}H_{0}^{\sigma^-}+B\sigma^{-}H_{0}^{\sigma^+}}{AH_{0}^{\sigma^-}+BH_{0}^{\sigma 
^+}}\right]^{2}.
\end{equation}
Now expressing (\ref{y6}) in terms of $\phi(t)$, using (\ref{p4}) and comparing with (\ref{y8}) we obtain that the relation 
$M_{0}^{2}=-H_{0}^{2}m^{2}$ is valid. This relation can be written in a more general manner as
\begin{equation}\label{y10}
\left.\left[\frac{\left(d\Gamma /d\psi\right)}{\Gamma}\right]^{2}\right|_{\psi =H_{0}^{-1}}=-H_{0}^{2}m^{2},
\end{equation}
which can be written as
\begin{equation}\label{mod2}
\left[\frac{A\sigma^{+}H_{0}^{\sigma^-}+B\sigma^{-}H_{0}^{\sigma^+}}{AH_{0}^{\sigma^-}+BH_{0}^{\sigma ^+}}\right]^{2}=-m^{2}.
\end{equation}
This expression for $\sigma^{+}=\sigma ^{-}=-3/2$ results satisfied for any $A$ and $B$ values and gives $m^{2}=-9/4$ which agrees 
with equation (\ref{sigma}).\\
The general solution of (\ref{y8}) has the form
\begin{equation}\label{y11}
\phi(t)=\phi _{0}\left[\alpha _{1}
e^{-\frac{H_0}{2}(3-\sqrt{9+4m^{2}})t}+\beta _{1}e^{-\frac{H_0}{2}(3+\sqrt{9+4m^{2}})t}\right],
\end{equation}
being $\alpha _{1}$ and $\beta _{1}$ dimensionless integration constants and $\phi _{0}$ some initial value for the scalar field 
$\phi(t)$. In the particular case when $\sigma^{+}=\sigma^{-}=-3/2$ the solution (\ref{y11})
becomes $\phi(t)=\phi _{0}[\alpha _{1}+\beta _{1}]e^{-(3/2)H_{0}t}$. These solutions, which
correspond to $m^2=-9/4$ (i.e., to $\sigma^{\pm}=-3/2$), are
decreasing with time as one expects for the expectation value of
the inflaton field during inflation. Furthermore, these values of
$\sigma^{\pm}$ avoid all possible divergences on $|\psi|
\rightarrow \infty $ on the general solutions (\ref{sol}).
However, the problem persists for $\psi =0$, for which the
solutions for $\varphi \propto \psi^{-3/2}$ would be divergents.
Of course, this problem can be solved by taking $\psi \neq 0$ in any foliation on the 5D vacuum. \\

In order to pass to a 4D induced phantom cosmology we follow the same treatment as in section II.
Thus, using form invariance transformations for a phantom scalar field $\bar\phi=i\phi$, the phantom scalar potential given by the 
transformation  $V_{ph}(\bar\phi)=\dot{\phi}^{2}/2+V_{eff}(\phi)$ acquires the form
\begin{equation}\label{y12}
V_{ph}(\bar\phi)=\frac{H_{0}^2}{2}\left[m^{2}-\frac{\left(\alpha _{1}\sigma^{+}e^{H_{0}\sigma^{+}t}+\beta 
_{1}\sigma^{-}e^{H_{0}\sigma^{-}t}\right)^{2}}{\left(\alpha _{1}e^{H_{0}\sigma ^{+}t}+\beta 
_{1}e^{H_{0}\sigma^{-}t}\right)^{2}}\right]\bar\phi\,^{2}.
\end{equation}
This phantom potential in the particular case previously considered $(\sigma^{+}=\sigma ^{-}=-3/2)$ becomes
\begin{equation}\label{mod3}
V_{ph}(\bar\phi)=-\frac{9}{4}H_{0}^{2}\bar{\phi}^{2}.
\end{equation}
Finally it must be noticed that the relation $a_{ph}(t)a(t)=1$ between the phantom scale factor $a_{ph}(t)=(1/a_0)e^{-2H_{0}t}$ 
and the scale factor $a(t)=a_{0}e^{2H_{0}t}$ is valid. It is important to mention that this duality relation is typical in phantom 
cosmologies \cite{Chimento3,Dab}.

\section{An induced 4D phantom cosmological scenario in presence of a decaying cosmological parameter}

 In the present section our goal is to investigate the possibility to derive an effective 4D phantom cosmological scenario from a 
5D apparent vacuum under the presence of a decaying cosmological parameter $\Lambda (t)$. In order to describe a 5D universe in 
apparent vacuum dominated by the presence of a cosmological parameter in a geometrical manner, let us to consider the 5D line 
element $dS^2=g_{AB} dx^A dx^B$\cite{lambda,Ponce}
\begin{equation}\label{s1}
dS^{2}=\psi ^{2}\frac{\Lambda (t)}{3}dt^{2}-\psi^{2}e^{2\int \sqrt{\Lambda(t)/3}\,dt}dr^{2}-d\psi ^{2},
\end{equation}
where $dr^{2}=\delta _{ij}dx^{i}dx^{j}$ is the euclidian line element in cartesian coordinates and $\psi$ is the space-like extra 
dimension. Adopting natural units ($\hbar =c=1$) the cosmological parameter $\Lambda (t)$ has units of $(length)^{-2}$.
The metric $h_{AB}$ in (\ref{s1}) is Riemann-flat so it satisfies $R^{A}\,_{BCD}=0$.\\

On this geometrical background we consider a purely kinetic scalar
field $\varphi$ whose dynamics is obtained from the 5D action
\begin{eqnarray}
{\cal S}&=&\int d^{4}x \,d\psi\,\sqrt{-\,^{(5)}\!g}\left[\frac{^{(5)}R}{16\pi G}-\frac{1}{2}g^{AB}\varphi _{,A}\varphi 
_{,B}\right] \nonumber \\
&\equiv &\int d^{4}x
\,d\psi\,\sqrt{-\,^{(5)}\!g}\left[\frac{^{(5)}R}{16\pi
G}-\left(\frac{3}{\psi^2\Lambda}\right)
\left[\frac{1}{2}\dot\varphi^2 -
\frac{1}{2a^2(t)}\left(\nabla_r\varphi\right)^2\right]
+\frac{1}{2}
\left(\frac{\partial\varphi}{\partial\psi}\right)^2\right],
\label{p2}
\end{eqnarray}
where $a(t)=\sqrt{(3/\Lambda)}\,e^{\int \sqrt{\Lambda/3}dt}$,
$^{(5)}R$ is the Ricci scalar (which in the case of the metric
(\ref{s1}) is null), $G=M^{-2}_p$ is the gravitational constant,
$M_p=1.2 \times 10^{19}\, {\rm GeV}$ denotes the Planckian mass.\\

The 5D scalar field dynamics is governed by
\begin{equation}\label{s2}
\ddot{\varphi}+\left[3\sqrt{\frac{\Lambda}{3}}-\frac{1}{2}\frac{\dot{\Lambda}}{\Lambda}\right]\,\dot{\varphi}-\frac{\Lambda}{3}\,e
^{-2 \int \sqrt{\Lambda/3}\,dt}\nabla _{r}^{2}\varphi -\frac{\Lambda}{3}\left[4\psi\frac{\partial\varphi}{\partial \psi}+\psi 
^{2}\frac{\partial ^2\varphi}{\partial \psi ^2}\right]=0.
\end{equation}
On hypersurfaces $\psi=\psi_0$ the effective action (\ref{p2}) can
be written as
\begin{equation}\label{act4}
{\cal S}_{eff}=\int d^{4}x \,\sqrt{-\,^{(4)}\!g}\left[\frac{^{(4)}R}{16\pi G}
-\frac{1}{2}g^{\mu\nu}\bar{\varphi} _{,\mu}\bar{\varphi} _{,\nu}+\frac{3}{\psi^2 \Lambda} V_{eff}(\bar{\varphi})\right],
\end{equation}
where $\mu$ and $\nu$ run from $0$ to $3$ and
\begin{equation}
^{(4)} {\cal R} = \left.\frac{12}{\psi^2}\right|_{\psi_0},
\end{equation}
is the effective 4D scalar curvature on the foliation
$\psi=\psi_0$.

 The classical effective 4D scalar potential after
using (\ref{sol}) adopts the form
\begin{equation}\label{s5}
V_{eff}(\bar{\varphi})=-\left(\frac{1}{2}\right)\frac{\psi _{0}^{2}\Lambda(t)}{3}\,g^{\psi\psi}\left.\left(\frac{\partial 
\varphi}{\partial \psi}\right)^{2}\right|_{\psi =\psi
_0}=\frac{\Lambda(t)}{3}\, \frac{\left[ A \sigma^+
\psi^{\sigma^+}_0 + B \sigma^-
\psi^{\sigma^-}_0\right]^2}{2\left(A \psi^{\sigma^+}_0 + B
\psi^{\sigma^-}_0\right)^2}\, \bar{\varphi}^2(\vec r,t).
\end{equation}

Now, let us to consider a decaying cosmological parameter of the
form $\Lambda(t)=3 p^{2}/t^{2}$, being $p$ a dimensionless
constant parameter. On a comoving coordinate system
($u^x=u^y=u^z=0$), the expectation value of $\varphi$ on the 3D
spatially flat, isotropic and homogeneous 3D volume ($x$,$y$,$z$)
is given by $\left<\bar{\varphi}(\vec r,t)\right>=\phi(t)$. With
this $\Lambda (t)$ equation (\ref{s2}), re-written in terms of
$\phi(t)$  becomes
\begin{equation}\label{s6}
\ddot{\phi}+\left[3\sqrt{\frac{\Lambda}{3}} -\frac{1}{2}
\frac{\dot\Lambda}{\Lambda}\right]\dot{\phi}+\frac{dV_{eff}(\phi)}{d\phi}
=0, \rightarrow
\ddot{\phi}+\frac{(3p+1)}{t}\,\dot{\phi}+\frac{M^{2}}{t^2}\,\phi
=0,
\end{equation}
where $(d V_{eff}(\phi)/ d\phi)\equiv \left.(d
V_{eff}(\varphi)/ d\varphi)\right|_{\varphi=\phi}$ and from
(\ref{p4}) we obtain that $M^2=-p^2 m^2$ is given by
\begin{equation}\label{9}
M^2 =  p^2 \frac{\left[A \sigma^+ \psi^{\sigma^+}_0 + B \sigma^-
\psi^{\sigma^-}_0\right]^2}{\left[A \psi^{\sigma^+}_0 + B
\psi^{\sigma^-}_0\right]^2}.
\end{equation}
From equations (\ref{s5}), (\ref{s6}) and (\ref{9}), we obtain
\begin{displaymath}
\left.\frac{\Lambda}{3} \psi^{2}_{0}
\left(\frac{d\Gamma/d\psi}{\Gamma}\right)^2\right|_{\psi_0} =
-\frac{\Lambda}{3} m^2,
\end{displaymath}
which for our example holds
\begin{equation}\label{10}
\frac{\left[A \sigma^+ \psi^{\sigma^+}_0 + B \sigma^-
\psi^{\sigma^-}_0\right]^2}{\left(A\psi^{\sigma^+}_0 + B
\psi^{\sigma^-}_0\right)^2} = -m^2.
\end{equation}
Notice that for the particular case
$\sigma^+=\sigma^-=\sigma=-3/2$, the eq. (\ref{10}) is fulfilled
for any $A$ and $B$ values.

The general solution of (\ref{s6}) is given by
\begin{equation}\label{s7}
\phi (t)=\phi_0 \left(\frac{t}{t_0}\right)^{-3p/2} \left[\alpha
\,\left(\frac{t}{t_0}\right)^{p\frac{\sqrt{9+4m^2}}{2}} +
\beta\,\left(\frac{t}{t_0}\right)^{-p
\frac{\sqrt{9+4m^2}}{2}}\right],
\end{equation}
where $\alpha$ and $\beta$ are dimensionless constants.  For
$\sigma^+ = \sigma^- = \sigma=-3/2$, we obtain the particular
solution $\phi(t) = (\alpha + \beta) \phi_0\, \left(t/
t_0\right)^{-3p/2}$, which decreases monotonically with time.
Now, implementing the form invariance transformations given in
\cite{Chimento1,Chimento3}, we can pass from our effective 4D cosmological scenario  with a real scalar field $\phi$ to a phantom
cosmological scenario with a phantom scalar field $\bar\phi =i\phi$. Thus, as it was shown in \cite{Chimento3}, the phantom scalar 
potential is given by $V_{ph}(\bar\phi)=\dot{\phi}^2/2+V_{eff}(\phi)$, where $\phi (t)$ and $V_{eff}(\phi)$ are determined by 
(\ref{s7}) and (\ref{s5}) respectively.
\begin{equation}\label{s8}
V_{ph}(\bar\phi)=\frac{1}{2}\left(\frac{\Lambda}{3}\right)\left[m^{2}-\frac{\left[\alpha
\sigma^+ \left(t/t_0\right)^{p\sigma^+} + \beta \sigma^-
\left(t/t_0\right)^{p\sigma^-} \right]^2}{ \left[\alpha
\left(t/t_0\right)^{p\sigma^+} + \beta
\left(t/t_0\right)^{p\sigma^-} \right]^2} \right] \bar\phi^2(t)
\end{equation}
For the particular case of $\sigma^+ =\sigma^-=\sigma=-3/2$, one
obtains
\begin{equation}\label{s9}
\left.V_{ph}(\bar\phi)\right|_{\sigma^+ =\sigma^-=\sigma=-3/2} =
-\frac{9}{4}\,\left(\frac{\Lambda}{3}\right)\,\bar\phi^2(t).
\end{equation}
Something interesting is that the phantom potential (\ref{s9}) has (for a constant $\Lambda =\Lambda _0$), algebraically the same 
form as the one derived in section II for a de-Sitter expansion (\ref{mod3}), if we consider $(\Lambda _0/3)$ playing the role of 
$H_{0}^{2}$. Of course, strictly  speaking both potentials are
different due to the time dependence of $\Lambda$.

\section{Final Comments}

Using form-invariance symmetry transformations in this letter we have derived 4D phantom cosmological scenarios induced from 5D 
vacuum gravity. The 5D vacuum gravity has been modeled using 5D Riemann-flat metrics. Our development has been focused on 
inflationary cosmology which is the subject of our main interest. As it is well-known during inflation the characteristic 
effective 4D equations of state are vacuum dominated, allowing the description of this period by means of a scalar field. In 
particular, we have studied two illustrative examples.

In the first one we passed from a 5D Riemann-flat metric to an effective 4D metric which describes a de-Sitter expansion. The 
dimensional reduction has been achieved by taking the foliation $\psi =H_{0}^{-1}$, being $H_{0}$ the constant Hubble parameter 
during inflation. In this case the effective 4D phantom potential obtained is quadratic on $\bar\phi$, with a negative constant 
effective squared mass.

In the second one we passed from 5D vacuum to an induced 4D vacuum dominated expansion driven by the inflaton field $\phi(t)$, in 
the presence of a decaying cosmological parameter $\Lambda \sim 1/t^2$. Now the dimensional reduction has been obtained by taking 
the foliation $\psi =\psi _0$. In this case the effective squared mass related to the effective 4D phantom field is linear with 
$\Lambda (t)$ and thereby in our case decreasing with time. Moreover for $\sigma^+ =\sigma^- = \sigma =-3/2$ the potential 
$V_{ph}(\bar\phi)$ results to be negative [see eq. (\ref{s9})], with a negative squared mass which has for a constant $\Lambda 
=\Lambda _0$. This has the same algebraic form as the one derived in the first example with $(\Lambda _0/3)$ playing the role of 
$H_{0}^{2}$.

An interesting aspect of these examples is that although both of them can describe 4D vacuum dominated expansions governed by the 
inflaton field, they are dynamically different due to the fact that their 4-velocities $u^{\alpha} = dx^{\alpha}/dS$ associated to 
their respective 4D induced metrics are different ($u^t$ is
constant in the first example, but time dependent in the second
one). Finally, since the symmetry
transformations here used are an effective 4D artifact connecting the expectation value of the effective 4D inflaton field
(which is a classical field) with the phantom field
$\bar\phi(t)$, we can say that the ``phantom
cosmologies '' here obtained from a 5D vacuum do not change the
original nature of the 5D bulk.\\

\vskip .2cm \centerline{\bf{Acknowledgements}} \vskip .2cm
\noindent
JEMA acknowledges CNPq-CLAF for financial support. MB
acknowledges CONICET and UNMdP (Argentina) for financial support.
\\

\end{document}